# A MULTI-OBJECTIVE PERSPECTIVE FOR OPERATOR SCHEDULING USING FINE-GRAINED DVS ARCHITECTURES

Rajdeep Mukherjee, Priyankar Ghosh, Pallab Dasgupta and Ajit Pal

Department of Computer Science and Engineering
Indian Institute of Technology Kharagpur
e-mail:{rajdeep, priyankar, pallab, apal}@cse.iitkgp.ernet.in

## ABSTRACT :

*The stringent power budget of fine grained power managed digital integrated circuits have driven chip designers to optimize power at the cost of area and delay, which were the traditional cost criteria for circuit optimization. The emerging scenario motivates us to revisit the classical operator scheduling problem under the availability of DVFS enabled functional units that can trade-off cycles with power. We study the design space defined due to this trade-off and present a branch-and-bound(B/B) algorithm to explore this state space and report the pareto-optimal front with respect to area and power. The scheduling also aims at maximum resource sharing and is able to attain sufficient area and power gains for complex benchmarks when timing constraints are relaxed by sufficient amount. Experimental results show that the algorithm that operates without any user constraint(area/power) is able to solve the problem for most available benchmarks, and the use of power budget or area budget constraints leads to significant performance gain.*

## 1. INTRODUCTION

The operator scheduling problem [12] is at the heart of every synthesis tool for digital integrated circuits. Given a data flow graph (DFG), where nodes represent operations and edges represent precedences between operations, the traditional goal of the operator scheduling problem has been to schedule the operations along clock cycles so as to minimize the total number of cycles (representing delay) and the total number of resources (representing area). There exists a significant body of literature which addresses this optimization problem [1, 2, 3, 12]. Another closely related problem is to determine the scheduling and binding for control and data flow intensive applications, also alternatively know as CDFG. Reducing the power requirement for CDFG has also been an important area of research [18, 19, 20, 21]. However in this paper we restrict our focus to DFG only and investigate the low-power scheduling of data-flow intensive applications under the availability of fine-grained power managed digital circuits under multiple user constraints.

DOI : 10.5121/vlsic.2013.4109                                                                           105

Over the last decade, power has become a major optimization criterion in all levels of circuit design, ranging from architectural level power management of large power domains to transistor level power management techniques like resizing and adaptive body biasing. While coarse grained power management at the level of architectural power domains (like processor cores and memory banks) has become ubiquitous, the use of fine grained power management at the level of individual functional units is a more recent phenomenon. Early work in this area centered around the use of multiple implementations of a given type of functional unit, each of which has the same functionality, but works on a specific voltage (Vdd) and requires a specific number of cycles. The operator scheduling problem under such Multi-Vdd resources as shown in Figure 1.1B has been studied in [4, 9, 10, 11]. These methods demonstrated the trade-off between area and power in the context of operator scheduling. Previous works [9, 10, 11] addressed low power scheduling of data-flow inten- sive circuits such as DSPs, which aims at maximum parallelism of operations. ILP based methods are explored in [14] where the objective was to minimize a scalar cost function comprising area, power and latency parameters. How- ever ILP based method often suffer from scalability issues, thereby making this technique inapplicable in the context of industrial size circuits. Several heuristic methods, such as those based on genetic algorithm [15] produce near-optimal results in a reasonable CPU time. A game theory-based ap- proach is presented in [17], which has considerable CPU running time for large VLSI systems. Traditionally resource and latency has been the most important optimization objectives. Existing scheduling techniques typically work with a single user constraint on one objective and try to optimize the other objective as much as possible.

Several advanced design techniques have been adopted for the implemen- tation of low power systems. All these techniques require additional hardware and software support for proper management of system power. Different de- sign abstraction levels incorporate different strategies for managing power. Typically, the power management can be done at system level, architecture, gate, circuit and the transistor level. All these design levels use some form of fine-grain or coarse-grain strategies of managing power, the simplest form of which involves turning off unused components of the chip when it is not func- tional or using lower operating voltage to perform non-critical operations. This requires additional power managed circuitry to ensure proper power- down of individual power domains as well as using multiple voltage/frequency pairs to implement low power designs.

As static power continues to rise due to the down-scaling trend of CMOS technology, fine-grained power management will have increasing significance in digital circuit design. More recently, fine-grained dynamic voltage scaling (FGDVS) technology is used to enable a single functional unit to work with different voltages and in correspondingly different number of cycles. It has been shown in [5, 6] that the use of DVS enabled resources reduces the area as compared to Multi-Vdd approaches. The authors of [5] demonstrate this gain, but do not present any algorithm for developing optimal schedules for operator graphs with such FGDVS enabled resources.

In our recent work [16], the classical operator scheduling problem during high level synthesis is revisited in the context of FGDVS enabled resources. We proposed a least cost branch and bound based bi-objective scheduling algorithm where resource, power are the non-commensurate objectives. Our algorithm works with an inherent strict latency constraint (equal to the length of the critical path), and additionally takes area or power budget as another constraint. The branch- and- bound formulation always guarantees to gener- ate the pareto-optimal solution for the given data-flow intensive applications under the presence of these multiple bounds. We also explored the extensions of the traditional list-based algorithm to handle different area or power budgets under a given latency constraint. It is observed that the algorithm fails to generate a solution in most cases due to multiple bounds. Moreover it fails to present the trade-off between area and power. A branch-and-bound formulation addresses this limitation and finds the solutions lying on the pareto-optimal front with respect to area and power. Since the problem is an instance of bi-objective optimization problem, the pruning-criterion is based on dominance, that is, a potential solution is discarded when it has exceeded all the criterion which are area and power of some previous solution.

In this work, we propose an extension to our existing approach and augment delay as another objective parameter along with area and power in the pareto-optimal solution frontier.The addition of the third dimension(area, power, delay) increases the solution space of the problem and hence the problem reduces to an instance of the multi-objective optimization problem. Again, the increase in delay has huge impact on the solution frontier comprising area and power. In this paper, we proposed a least cost branch and bound based multi-objective scheduling and binding algorithm where resource, power and delay are the non-commensurate objectives.

We present results on standard data-flow intensive benchmarks from DSP domain. Experimental results reveal that the algorithm is able to report the solutions on the pareto-optimal front within feasible limits of time, which is adequate for practical purposes, since operator scheduling is likely to be done only once in a given design. In this variant of the operator scheduling problem, it is extremely hard to predict the trade-off between the area and power. Therefore systematic traversal of the state space can yield the non- dominated solutions in sequences which may be non-monotonic with respect to both criteria. In such state spaces, constraints on one or more dimensions typically prove to be very useful in pruning the state space.

The paper is organized as follows. Section 1.2 discusses the FGVDS framework. Section 1.3 presents an example motivating the problem of operator scheduling with FGDVS enabled resources. Section 1.4 presents the definitions. Section 1.5 formally defines the problem. Section 1.6 presents the overhead of FGDVS architecture. Section 1.7 presents the list-based algorithm and branch-and-bound algorithm. Section 1.8 demonstrates the experimental results on high-level synthesis benchmarks. Section 1.9 presents our conclusions on the proposed approach.

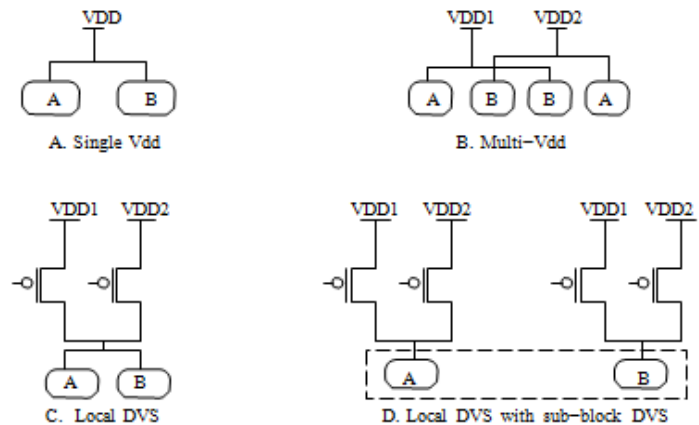

Fig. 1.1 CMOS architectures

## 2. FINE - GRAINED DVS FRAMEWORK

Figure 1.1C and Figure 1.1D shows FGDVS enabled resources that implements DVS with local header switches that are inserted down to the sub- block level. The fine-grained header switches add temporal granularity to switch between different processing rates and enables a functional unit to be reused with a different operating voltage, instead of permanent assignment of voltage to components as done in Multi-Vdd enabled functional units. This architecture combines the benefits of Multi-Vdd, fine-grained header switches and dynamic voltage scaling to achieve a wide range of processing rates as well as maximum energy savings.

FGDVS architecture is most beneficial for systems that implement time- wise mutually exclusive functions on the same hardware. The benefits also increase for DFGs with heterogeneous operations compared to Multi-Vdd where different functions have to be implemented with different components at each voltage level. When the available slack is sufficient enough such that it allows heavy energy consumer operations of the DFG to be scheduled at lower voltages, then fine-grained DVS helps to achieve sufficient gain both in terms of area and power parameters.

The fine-grained power management architecture is suitable for mitigating the increase in energy loss due to leakage. The functional modules that are unused during execution of a given DFG consume leakage energy for single-Vdd and Multi-Vdd approach but FGDVS architecture with component-level header switches could switch-off unused components using power gates leading to minimal leakage during standby or sleep mode. Further, using FGDVS, the total number of resource instances of each type required to schedule a operator graph is often less than the other two approaches, which in turn reduces leakage power as discussed in Section 1.4.1 as well as makes it more area efficient despite the presence of additional supply rails.

## 3. MOTIVATING EXAMPLE

Figure 1.2A presents a simple data flow graph (DFG) and the corresponding three different schedules, B,C and D. All three schedules complete within the critical time of 5 control steps, where each control step (c-step) corresponds to one clock cycle. We are given resources that operate at three different voltages, 1.0V, 0.78V, 0.68V and requires 1 cycle, 2 cycles, and 3 cycles respectively. Schedule B uses the Single-Vdd approach where each resource can operate only at a fixed voltage (1V) and hence requires the maximum power. Schedule D uses two multipliers when static multi-Vdd multipliers of 1V and 0.78V are used.

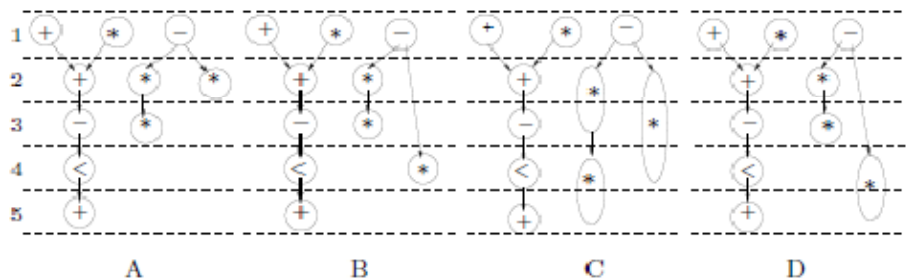

Fig. 1.2 DFG and its corresponding Schedules

| Architecture | Area | Power(in mW) | | | Total Power |
|---|---|---|---|---|---|
| | | Dynamic | Leakage | Switching | |
| Single-Vdd | 4 | 105.44 | 8.12 | φ | 113.56 |
| Multi-Vdd | 4 | 105.44 | 8.12 | φ | 113.56 |
| | 5 | 80.54 | 9.94 | φ | 90.48 |
| | 6 | 64.28 | 15.65 | φ | 79.93 |
| FGDVS | 4 | 92.96 | 2.16 | 1.56 | 96.68 |
| | 5 | 64.55 | 3.45 | 1.56 | 69.56 |

Table 1.1 Non-dominated (area, power) tuples for the DFG in Figure 1.2A

On the other hand only one DVS enabled multiplier suffices for all the multiplication operations of Schedule D thereby reducing area without com- promising power as compared to the static Multi-Vdd approach. Schedule C requires three multipliers for static Multi-Vdd approach (each operating at different voltages) compared to two multipliers required using fine-grained DVS architecture. This happens because FGDVS allows time-wise mutually exclusive functions to be scheduled on the same hardware thereby reducing area and in turn leakage power, but contributing to Vdd-Switching power since in both schedule C and D, a multiplier operating at 1.0V is reused by switching it to 0.78V in consecutive control steps as shown in Table 1.1. The table also shows the non-dominated solutions for the Multi-Vdd and DVS approaches. For the classical single-Vdd approach, only one solution, namely the one which minimizes area is reported.

## 4. DEFINITIONS

A data flow graph (DFG), G = (V, E), is a directed acyclic graph, where every node $v_i \in V$ represents an operation and an edge $e \in E$ from node vi to vj represents the precedence constraint between $v_i$ and $v_j$.

The mobility of a node vi is the difference between its as late as possible (ALAP) schedule time and its as soon as possible (ASAP) schedule time [3,12]. The ASAP and ALAP times are computed assuming that the latency is equal to the length of the critical path.

A schedule s is a mapping from the nodes to a pair denoted by :: (allocation time, duration). Allocation time is the control step in which a node is scheduled. Duration of a node is defined as the number of control steps that can be allocated to a node such that this value never exceeds the difference between the ALAP time and present allocation time of a node. The choice of duration of an operation reflects the choice of the operating voltage of the resource and that choice may be restricted by the mentioned difference between the ALAP time and present allocation time of a node. The cost of a schedule s, denoted by Cost(s), is expressed in terms of the pair (area, power, time).

Cost $c_1 = (a_1, p_1)$ dominates cost $c_2 = (a_2, p_2)$, denoted by $c_1 \prec c_2$, iff : (a) $a_1 \leq a_2$ and $p_1 \leq p_2$, and (b) $a_1 < a_2$ or $p_1 < p_2$.

International Journal of VLSI design & Communication Systems (VLSICS) Vol.4, No.1, February 2013

A multi-objective optimization problem aims at searching for solutions where attempt to improve an objective further, degrades the quality of other objectives. Such tentative solution is called non-dominated, Pareto optimal, or Pareto efficient. The goal of a multi-objective optimization problem is to find such non-dominated solutions, and quantifying the trade-offs in satisfying the different objectives.

## 4.1 POWER MODEL

There are three major sources of power consumption for fine-grained DVS enabled functional modules::

A. Dynamic Power :: The dynamic power consumed by a functional module is given by: $P_{dyn} = \alpha.V^2_{dd}.C_{load}.f_{switch}$, where $\alpha$ is the switching activity, $C_{load}$ is the load capacitance and $f_{switch}$ is the activity of the module. The product term $C_{load}.f_{switch}$ in the dynamic power equation assumed to be constant for each functional module following previous literature [10]. This assumption is used in high-level synthesis because the values of these parameters are not known at this stage of synthesis and depends on the interconnection pattern, placement of the operator in the data-path, and the input sequence.

B. Leakage Power :: The sub-threshold leakage power is given by [8]: $P_{lkg} = Ae^{(V_{gs} - V_{th} - V_{off})/mVT}(1 - e^{-V_{ds}/VT}).V_{dd}$, the fine-grained DVS architecture uses PMOS power switches, or headers that uses a small set of discrete voltage and frequency pairs to connect to the DVS functional modules and approximate a broad range of energy/performance points. During oper- ator scheduling, when a device is not used in a particular control step, then it is power-gated with a transistor of high threshold voltage. Thus, header switches allow systems to switch-off leakage current when a circuit block is not used. In contrast, leakage power is significant in Multi-Vdd architecture since it has larger resource requirement compared to FGDVS where fewer components are active at any time for the same schedule of the DFG.

C. Vdd-Switching Power :: Further, the local DVS achieves fine-grained power control by dithering the header switch on or off and allows a circuit to switch between different Vdd's, which contributes to Vdd-Switching Power. The Vdd-Switching power is given by [6]: $P_{switch} = (I_{ddh} V_{ddh} + I_{ddl}V_{ddl})/2 - P_{dyn}$, where $P_{dyn}$ is the dynamic power, $I_{ddh}$, $I_{ddl}$ is the average current through the high ($V_{ddh}$) and low voltage($V_{ddl}$) respectively. The algorithm computes an overestimate of the switching power by accounting switching for all possible scenarios except when for the current time cycle and for a particular operation there is a free resource which was allocated earlier with the same duration. Thus, the algorithm considers the $V_{dd}$-switching power that comes from changing the supply voltage. It has been shown in [6] that smaller header switches has lower voltage switching overhead but higher voltage switching delay.

With this assumption, the power dimensions considering the dynamic, leakage and $V_{dd}$-Switching power, and the voltage and delay parameters in our optimization problem has been borrowed from [6] and the total power of a schedule is computed by the equation: $L (P_{dyn} + P_{lkg} + P_{switch})$ over all modules along all control steps. These power numbers has been obtained for 90 nm CMOS technology. Our algorithms are generic and can work with power numbers from other technologies as well.

## 5. PROBLEM DEFINITION

Given a data-flow graph, a latency bound, and an area or power budget, our objective is to generate pareto-optimal schedule of (area, power, time) tuples satisfying these constraints using fine-grained DVS enabled functional modules. It may be noted that throughout this paper we assume strict or relaxed timing constraints to perform the scheduling operations. The strict timing constraint is such that that the length of all schedules are within the number of cycles in the critical path of the DFG, and in such case the mobility of the critical path operations are zero. In other words, the scope of the optimization is limited to choosing the appropriate resources for the operations on both critical and non-critical paths and scheduling them within the given timing bounds.

In the classical operator scheduling problem, the possible number of choices for scheduling an operation is upper bounded by its mobility. The product of the mobilities of all the critical and non-critical operations is an upper bound on the state space of the operator scheduling problem, though the actual state space is smaller, since scheduling one operation can reduce the mobility of many others.

In the case of scheduling with static Multi-$V_{dd}$ resources, the choice of the operating voltage (and consequently the number of cycles) for each operation is another variable. Hence the state space of the operator scheduling problem is multiplied by this factor. The situation is similar for scheduling with DVS enabled resources, except that the reuse of resources entails a different function for computing the area. Nevertheless, the state spaces of the two problems (with Multi-Vdd and DVS) have similar structure and a branch- and-bound on one can be used to traverse the other as well. We now present a branch-and-bound and a list-based algorithm for this problem.

Further, the impact of strict timing constraint and relaxed timing constraint is studied in this paper for schedules generated using fine-grained DVS architectures. More the relaxation of timing bound, larger the increase in the state space of the problem and it takes more time for the B/B formulation to traverse the solution space and generate the pareto-optimal frontier. But the advantage of relaxing the timing constraint is that the power consumption values may decrease at the same rate due to availability of slack during scheduling.

## 6. OVERHEAD OF FGDVS ARCHITECTURE

The main overheads of FGDVS that influence scheduling algorithms on this architecture are $V_{dd}$-Switching Power and $V_{dd}$-Switching delay. When a volt- age switch occurs, the time required to charge or discharge gates of header switches results in Switching delay and the amount of charge delivered to the switches and virtual rails results in Switching energy overhead. The scheduling decision is made such that the power required to schedule an operation of a DFG at low voltage plus the Vdd-Switching power for switching the unit down and up again must be less than the power consumed for execution of that operation at highest voltage. The availability of DVS enabled functional units have a significant impact on the size of the state space for the operator scheduling problem. The area overhead of this architecture comes from additional header switches, and level converters. It has been shown in [5, 6] that both of these overheads are proportional to the size of the header switches.

# 7. PROPOSED ALGORITHMS

Our proposed algorithms always work with a strict or relaxed latency con- straint. Additionally, it takes an user constraint such as area or power bud- get. In this paper, we propose the following algorithms : 1) An extension of traditional List-based scheduling algorithm with the ability to handle area or power budgets, under the given latency constraint and 2) A Branch and Bound(B/B) formulation for exploring the pareto-optimal front.

We also study a variation of the B/B algorithm to handle different area or power budgets under the given latency constraint. A simple B/B formulation with only latency constraint and no user constraint, is also implemented with different types of functional units such as ::

1) Single-$V_{dd}$, 2) Multi-$V_{dd}$, 3) Fine-grained DVS and a comparative study is presented.

## 7.1 A LIST-BASED ALGORITHM

List scheduling is one of the most popular scheduling strategies under resource constraints. We extended the traditional version of the algorithm to work with a strict latency bound, that is when the timing constraint is equal to the length of the critical path. The list of ready nodes to be scheduled in the present time step is assigned to its available duration values based on some priority in order to satisfy the given area budget as well as the latency bound. However, the List scheduling algorithm in absence of resource constraint behaves similar to ASAP scheduling. In this case, the algorithm works with a latency bound and different power budgets as an user constraint. The nodes in the critical path have zero mobility and are already scheduled in their corresponding time-step. The algorithm operates on a topologically ordered list, L, of non-critical nodes. For every node, $v_i$, in the list, the algorithm computes the corresponding allocation time and the available duration.

An operation in the ready list is allocated either its maximum or minimum available duration (which are the two different priority functions) such that the given constraints are satisfied. A node belonging to a non-critical path is chosen only after scheduling every parent of that node. However, due to the presence of multiple user constraints, the algorithm may fail to generate any schedule for a DFG under different area or power budgets, as shown in Table 1.2, 1.3.

## 7.2 PROPOSED BRANCH AND BOUND ALGORITHM

Our proposed algorithm uses branch-and-bound using the dominance relation to compute the pareto-optimal frontier of schedules in terms of the area and power attributes. The algorithm maintains the set o f non-dominated schedules found so far and prunes the search path whenever the partial solution is dominated by some previous solution. We present different versions of branch-and-bound, all of which has an inherent latency bound.

International Journal of VLSI design & Communication Systems (VLSICS) Vol.4, No.1, February 2013**Algorithm 1: Least Cost Branch and Bound Method**

```
input  : A precedence constrained DFG, G = (V, E), a area/power budget B, and a
         latency constraint
output : Pareto-optimal schedules of the data flow graph
1  foreach node v₁ ∈ V do
2  |    Compute the ASAP, ALAP, mobility of vᵢ;
3  end
4  Construct a list, L, of critical and non critical nodes;
5  Sort L according to the topological order;
6  Construct a set, S, of schedules, which is initially empty;
7  ScheduleBB(1);
```

**Procedure ScheduleBB(nodeIndex)**

```
1   if BoundExceeded(nodeIndex) then
2   |    return;
3
4   if nodeIndex > number of total nodes then
5   |    UpdateSchduleList();
6   |    return;
7   end
8   vᵢ ← L[nodeIndex];
9   tᵢ ← The earliest time instant vᵢ can be scheduled;
10  for t ← tᵢ to ALAP(vᵢ) do
11  |    δ ← The maximum value of the duration that can be assigned to vᵢ. given
        |        that vᵢ starts at t;
12  |    for d ← 0 to δ do
13  |    |    Schedule vᵢ at time instant t with duration δ;
14  |    |    ScheduleBB(nodeIndex + 1);
15  |    end
16  end
```

The algorithm uses an optional user constraint, power budget constraint, P or an area budget constraint, A and finds all pareto-optimal solutions satisfying the given power budget, area budget respectively. Another version of B/B algorithm that runs without any user constraints generates entire pareto-optimal schedules for a given data-flow graph. The authors of [7] also show the existence of the pareto-optimal frontier, but no algorithm was presented to generate the front.

We use the proposed approach to generate the pareto-optimal schedules for both DVS and Multi-Vdd approaches by applying appropriate functions for computing the cost of the schedule. Since the cost function is used only for pruning purposes, our approach serves as a general framework for exploring the set of non-dominated schedules for a wide variety of cost functions.

```
Function BoundExceeded(nodeIndex)
1   V_cur ← φ;
2   for i ← 0 to nodeIndex − 1 do
3   |   V_cur ← V_cur ∪ L[i];
4   end
5   Compute the cost, c = (a, p), of the partial schedule containing the nodes in V_cur;
6   if cost of partial schedule exceeds budget B then
7   |   return true;
8   else
9   |   foreach schedule s ∈ S do
10  |   |   if Cost(s) < c then
11  |   |   |   return true;
12  |   |   end
13  |   end
14  end
15  return false;
```

```
Procedure UpdateScheduleList
1   s_cur ← The current schedule;
2   c ← Cost(s_cur);
3   foreach schedule s ∈ S do
4   |   if c < Cost(s) then
5   |   |   Remove s from S;
6   |   end
7   |   Add s_cur to S;
8   end
```

Given a DFG, our algorithm first computes the ASAP, ALAP and the mobility values for all nodes in the graph under the given timing constraint. A topologically ordered list, L, of non-critical nodes is used to implement the branch-and-bound search. The search is done recursively by computing all possible allocation times and duration times for every node, $v_i$, in the list of critical and non-critical nodes, L, and then recursively scheduling the remaining nodes in L for every allocation time and duration of $v_i$. The search (Procedure ScheduleBB) schedules the nodes (by assigning the allocation time and duration) in the topological order. Therefore, a node in the graph is chosen only after scheduling every parent of that node.

Procedure BoundExceeded handles the decision of pruning depending on the cost of the partial schedule generated so far. If the cost tuple of the partial schedule is dominated by any previous solution, or if the input area or power constraint is violated, then that branch is pruned and the alternative assignments of allocation time and duration of the previous node in L are tried. The BoundExceeded function uses a small heuristic that prunes a partial schedule when the corresponding area or power value exceeds the given budget B.

For generating the schedules with FGDVS and Multi-Vdd approaches, different cost functions are used in the BoundExceeded function. The estimated area for DVS approach is

114

computed by summing up the maximum number of resources of each type used over all time-steps. For the Multi-Vdd approach, for each resource type we need to distinguish between the instances using different operating voltages and count the number of resources accordingly.

It is important to note that during the running of the algorithm, a solution in L may become dominated by a new solution and may therefore have to be removed from L. This check is performed by the procedure, UpdateScheduleList.

An important observation is that the order of exercising the scheduling choices does not guarantee that successive solutions will be monotonic in area or power. This means that until the algorithm terminates we cannot claim that any of the solutions in L actually belong to the pareto-optimal front. The consequence of this observation is that the search must run to completion should we wish to guarantee that any chosen solution is non-dominated.

### **7.3 PROOF OF CORRECTNESS**

Theorem 1. For a given power constraint $P$, every pareto-optimal schedule s having power number less than or equal to $P$, is generated by Algorithm 1. Proof. Suppose Algorithm 1 is run on a precedence constrained DFG, $G = (V, E)$, with a power constraint $P$. For the purpose of contradiction, let us assume that there exists a pareto-optimal solution, s, with cost $c = (a, p)$ and $p < P$, such that s is not generated by Algorithm 1.

For each $v_i \in V$, the schedule s assigns an allocation time and a duration to $v_i$. Since a tight timing constraint is used, i.e., latency being equal to critical path length, the nodes in the critical path have a fixed allocation time and duration for all schedules. Also consider the fact that every node, $v_i$, in a non-critical path of G, must satisfy the mobility condition, which states that (a) start time of $v_i$, $\alpha(v_i) \geq$ ASAP value of $v_i$, and (b) $\alpha(v_i) + \delta \leq$ ALAP value of $v_i$, where $\delta$ is the assigned duration of $v_i$ in s.

From the description of Algorithm 1 it follows that

1. Algorithm 1 schedules the non-critical nodes in topological order. Therefore, a node belonging to a non-critical path is scheduled after scheduling every parent of that node.
2. Procedure ScheduleBB explores all possible values of allocation time and duration for a non-critical node, for a particular assignment of allocation time and duration of its parent nodes
3. Function BoundExceeded prunes a branch when (a) either the cost of current partial schedule exceeds the given power budget, or (b) there exists a solution which is already generated and has a cost that dominates the cost of the current partial schedule.

Consider the solutions to which the pruning condition is not applicable. The schedule corresponding to the critical nodes is fixed and the schedule for the non-critical node will satisfy the mobility condition. Therefore these schedules will be generated by Algorithm 1. Since the pruning condition is not applicable to solution s, the assignment of allocation time and duration will also be explored by Algorithm 1 – a contradiction which proves the statement of Theorem 1.

International Journal of VLSI design & Communication Systems (VLSICS) Vol.4, No.1, February 2013## 8. EXPERIMENTAL RESULT

For our experimentation, we run our algorithms on a set of high-level synthe- sis benchmark circuits [13] from the Digital Signal Processing (DSP) domain. Following the set up in [5, 6], we consider functional units operating in three supply voltages, namely, 1.0V, 0.78V and 0.68V. It is worth noting that the additional level shifters required for approaches that use multiple voltages do not incur a significant overhead [10]. The power, time values in the result section are in milliWatt(mW), seconds respectively. The experimentation has been carried out on a 3.00 GHz Intel Core 2 Duo machine with 4 GB RAM.

| Circuit | Area Budget (Mul, Add, Comp) | using FGDVS enabled resources | | | | | |
|---|---|---|---|---|---|---|---|
| | | List-based (Area, Power(mW)) tuples | | | | B/B First Solutions | Time (Sec) |
| | | Priority1 | Time(Sec) | Priority2 | Time(sec) | | |
| Diffeq | (3, 2, 1) | (6, 122.3) | 0.02 | (6, 152.5) | 0.03 | (6, 122.3) | 0.08 |
| | (3, 1, 1) | φ | 0.04* | φ | 0.02* | (5, 147.2) | 0.12 |
| IIR | (4, 2, 1) | φ | 0.03* | (7, 195.8) | 0.04 | (7, 128.2) | 0.73 |
| | (3, 2, 1) | φ | 0.03* | (6, 195.8) | 0.07 | (6, 144.58) | 0.08 |
| FIR | (3, 1, 0) | (4, 145.1) | 0.45 | (4, 284.3) | 0.41 | (4, 145.1) | 2.92 |
| | (2, 1, 0) | (3, 184.7) | 1.33 | (3, 284.3) | 1.35 | (3, 184.7) | 3.77 |
| Volterra | (7, 4, 0) | φ | 3.56* | (11, 414.2) | 3.78 | (11, 392.2) | 10.38 |
| | (7, 3, 0) | φ | 2.34* | (10, 414.2) | 2.54 | (10, 401.77) | 11.34 |
| AR-Lattice | (6, 3, 0) | (9, 323.8) | 0.55 | (9, 399.2) | 0.64 | (9, 323.8) | 1.094 |
| Filter | (4, 2, 0) | φ | 0.24* | (6, 399.2) | 0.78 | (6, 349.4) | 1.568 |
| EWF | (3, 4, 3) | (10, 308.9) | 1.23 | (10, 308.9) | 1.56 | (10, 308.9) | 3.923 |
| | (2, 4, 2) | φ | 1.33* | φ | 1.43* | (8, 304.3) | 4.231 |

Table 1.2 Results demonstrating the working of B/B and Extended List-based Approach under different Area Budgets and strict timing constraint

Table 1.2 shows that List-scheduling fails to find a solution for most bench- marks when an operation is extracted out of the ready list and is assigned in its highest available duration value as shown in Column 3 (priority 1). We use φ to denote the entries corresponding to the cases for which no solutions could be found for the given budget (the runtime is highlighted using '∗'). However, when operations in the ready list are assigned with their minimum.

116

| Circuit | Power Budget | using FGDVS enabled resources | | | | | |
|---|---|---|---|---|---|---|---|
| | | List-based (Area, Power(mW)) tuples | | | | B/B First Solutions | Time (Sec) |
| | | Priority2 | Time(Sec) | Priority1 | Time(sec) | | |
| Diffeq | 190 | (6, 152.5) | 0.02 | (7, 118.5) | 0.05 | (5, 147.2) | 0.06 |
| | 115 | φ | 0.03* | φ | 0.07* | (7, 113.4) | 0.18 |
| IIR | 230 | (10, 195.8) | 0.02 | (11, 115.3) | 0.05 | (7, 148.75) | 0.73 |
| | 120 | φ | 0.03* | (11, 115.25) | 0.07 | (7, 119.12) | 0.06 |
| FIR | 300 | (12, 284.31) | 0.42 | (12, 141.14) | 0.34 | (12, 284.31) | 2.73 |
| | 150 | φ | 1.23* | (12, 141.14) | 1.26 | (6, 148.92) | 4.56 |
| Volterra | 455 | (15, 414.22) | 1.45 | (15, 302.17) | 2.33 | (15, 414.22) | 10.45 |
| | 305 | φ | 2.23* | (15, 302.17) | 2.14 | (15, 302.17) | 12.36 |
| AR-lattice Filter | 425 | (12, 399.16) | 0.44 | (12, 323.26) | 0.35 | (12, 399.16) | 2.34 |
| | 330 | φ | 0.18* | (12, 323.26) | 0.65 | (10, 323.26) | 1.45 |
| EWF | 370 | (8, 329.53) | 1.21 | (11, 303.72) | 1.45 | (8, 324.8) | 3.78 |
| | 280 | φ | 1.19* | φ | 1.21* | (12, 279.12) | 4.67 |

Table 1.3 Results demonstrating the working of B/B and Extended List-based Approach under different Power Budgets and strict timing constraint

| Circuit | No. of nodes | Single VDD (Area, Power) | Multi-VDD (Area, Power(mW)) | Time | Fine-grained DVS (Area, Power(mW)) | Time |
|---|---|---|---|---|---|---|
| Diffeq | 11 | (5, 188) | (5, 157),(6, 136),(7, 125) | 0.09 | (5, 143),(6, 118),(7, 113) | 0.65 |
| IIR | 16 | (6, 227) | (6, 186),(7, 163),(8, 153) (9, 144) | 1.23 | (6, 145),(7, 128),(8, 116) (9, 115) | 5.13 |
| FIR | 21 | (3, 299) | (3, 299),(4, 240),(5, 194) (6, 189) | 2839.27 | (3, 185),(4, 145),(5, 141) | 2711.87 |
| Volterra | 28 | (10, 451) | (10, 431),(11, 420),(12, 408) (13, 384),(14, 374),(15, 350) (16, 339) | 69.19 | (10, 402),(11, 389),(12, 367) (13, 347),(14, 322),(15, 302) | 92.18 |
| Lattice Filter | 28 | (6, 422) | (6, 422),(7, 399),(8, 377) (9, 369) | 10.89 | (6, 349), (7, 339),(8, 332) (9, 323),(10, 323) | 16.63 |
| EWF | 37 | (8, 367) | (8, 367),(9, 346),(10, 342) (11, 334),(12, 332) | 17.28 | (8, 299),(9, 292),(10, 285) (11, 280),(12, 277) | 13.96 |
| DCT | 42 | (10, 1050) | (10, 1050),(11, 983), (12, 938) (14, 900),(15, 884),(13, 916) (16, 868),(17, 852),(18, 836) (20, 807),(22, 794) | 9070.06 | (10, 1018),(11, 935),(12, 890) (14, 842),(15, 810),(13, 858) (16, 807),(18, 794),(19, 823) | 8564.66 |

Table 1.4 Pareto-Optimal (Area, Power) frontier for different DSP benchmarks using simple B/B approach under strict timing constraint available duration, then List scheduling generates solution for most bench- marks. Column 5 (Priority 2) of Table 1.2 demonstrates that the algorithm generates a solution within the given area budget, but there is no improvement in corresponding power numbers even for schedule with different area budgets. This happens due to greedy way of assigning priority to the nodes in the ready list and scheduling them within a latency bound. This warrants exhaustively searching the state space of the data-flow graph to generate schedule with better area, power values within feasible limits of time. Our Branch and Bound formulation(B/B) always guarantees to terminate with a solution given a area budget. Column 7 reports the first solution under different area budgets. The first solution obtained is not necessarily a non- dominated one, and one needs to run the algorithm to completion in order to find non-dominated solutions under the area budget. The value of the result shown in Table 1.2 is in determining quickly whether any solution exists under the given area budget – a requirement which may be a part of a more elaborate design space exploration.

List scheduling algorithm is a generalization of the ASAP scheduling in absence of resource constraint. So, we run the algorithm with power budget as well along with a timing constraint equal to the length of critical path. Column-3, Column-5 of Table 1.3 demonstrates that the algorithm even fails to generate a solution for few benchmarks within the given power budget, irrespective of the order in which the nodes in the ready list are scheduled based on their duration values. This happens because the nodes in the ready list are always assigned in their ASAP time-step with available duration and so the possibility of allocating other nodes within its mobility is further re- stricted. Thus, the algorithm fails to generate a solution within the given time and power budget. However, the branch and bound algorithm always ensures a solution for the same power and time budget. Column 7 of Table 1.3 reports the first solution found by B/B algorithm with the given power budget and the runtime to find that solution. The first solutions are the immediate power number of the schedules satisfying the given constraint, and the generated first solution is always orders of magnitude faster compared to the time re- quired by the same B/B algorithm when run in an environment with no user constraint. The power budgets were chosen between two crude estimates. A crude lower bound was obtained by considering the power of the critical path operators added with the lowest power option for all non-critical nodes. A crude upper bound was obtained by using the power of the first schedule (that is, the ASAP schedule). The results show a remarkable difference with the run-times shown in Table 1.4 for finding the entire pareto-optimal front. However, the simple Branch and Bound always provides optimal schedules for different benchmarks.

Table 1.4 shows the experimental results for running our algorithm with an inherent latency bound and without any user constraints on the benchmarks in [13]. Column-2 shows the number of nodes in the operator precedence DFG. Column-3, Column-4 and Column-5 report the non-dominated solutions obtained using the Single-$V_{dd}$, Multi-$V_{dd}$ and DVS approaches respectively. Additionally in Column-4 and Column-5, we report the run-times to compute the entire pareto-optimal set of solutions. The results presented in Table 1.4 demonstrate several interesting aspects of the methodology and our algorithm. Firstly we observe that the pareto-optimal front of DVS dominates the pareto-optimal front of the Multi-$V_{dd}$ approach. Previous researchers illustrated cases where a DVS solution dominated a Multi-$V_{dd}$ solution [5], but our results take it further to show that almost every pareto-optimal solution of Multi-$V_{dd}$ is dominated by a DVS solution. Figure 1.3, Figure 1.9 shows the Multi-$V_{dd}$ and DVS pareto-optimal fronts for the EWF and DCT benchmarks graphically.

The run-times of our algorithm demonstrate that for the scale of the problem at hand, pure branch-and-bound works well within feasible limits of time. We also executed the algorithm to obtain DVS schedules under different power budgets.

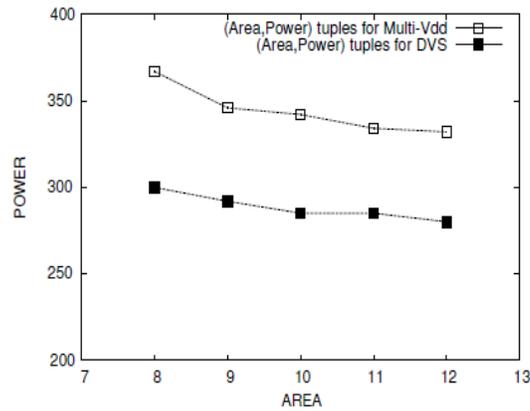

Fig. 1.3 Pareto-Optimal (area, power) frontier using Multi-Vdd and DVS for EWF

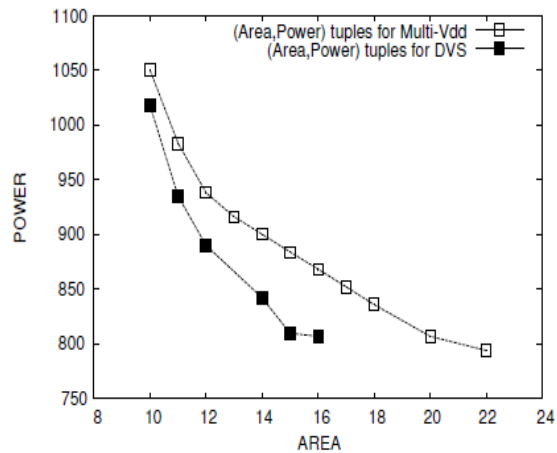

Fig. 1.4 Pareto-Optimal (area, power) frontier using Multi-Vdd and DVS for DCT benchmark circuit

Figure 1.5 shows the change in maximum as well as minimum area values when the circuits are run with different allowable slack, expressed in terms of number of additional control steps(k). It is observed that the rate of decrease in minimum area values (number of resources of each type used) generated during the pareto-solution frontier is more compared the maximum area values lying in the pareto-optimal solution frontier as the allowable slack increases. Here, k = 0 denotes strict timing constraint i.e., when the timing bound for a schedule is equal to the length of the critical path. Thus, the relaxed timing constraint generates schedules with lower minimum area requirements.

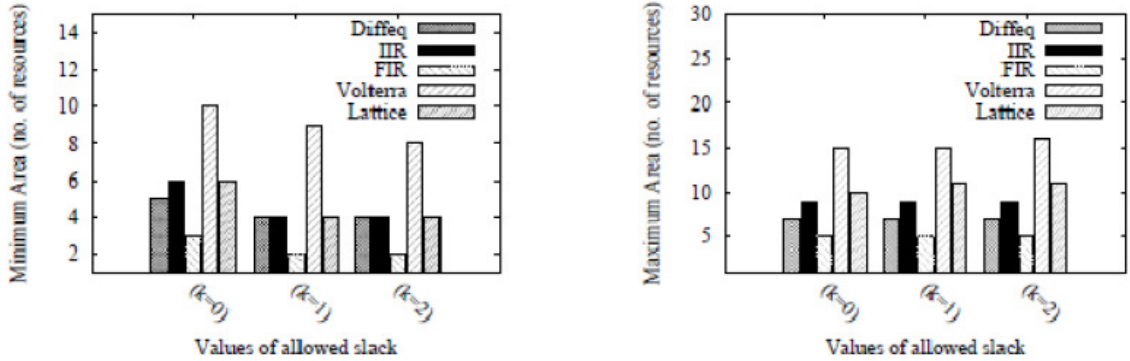

Fig. 1.5 Variation of Maximum Area and Minimum Area values with increase in slack for different benchmarks

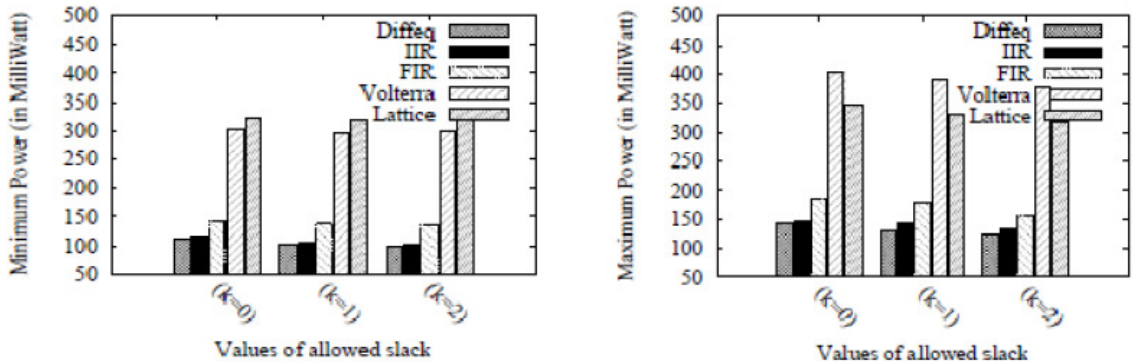

Fig. 1.6 Variation of Maximum Power and Minimum Power values with increase in slack for different benchmarks

Figure 1.6 shows the change in maximum as well as minimum power values when the circuits are run with different allowable slack, expressed in terms of number of additional control steps(k). It is observed that the rate of decrease in maximum power value (considering dynamic power, leakage power and switching power) generated during the pareto-solution frontier construction is more compared to the minimum power value lying in the pareto-optimal solution frontier as the allowable slack increases. Thus, the increase in slack reduces the power of the DFG benchmarks by considerable amount. This happens because the allowable slack increases the mobility for an operator, and thus it can be bind to functional units consuming lower voltage and hence lower power.

## 9. CONCLUSION

This paper shows that with existing computational resources, a simple branch-and-bound technique is able to compute, within feasible limits of time, the entire pareto-optimal set of schedules for the operator scheduling problem with DVS-enabled resources. It is also shown that the same algorithm can be used to quickly evaluate whether a schedule exists under a given power budget and area budget. The branch-and-bound formulation is generic and hence allows us to evaluate and compare operator schedules under different low power techniques, such as Multi-Vdd and FGDVS, by simply changing the cost function. A comparative study is also presented between the B/B and list-based technique for different DFG benchmarks under different area and power budgets. The cost function for modeling power in FGDVS architecture considers the dynamic, leakage and $V_{dd}$-switching power. The use of relaxed timing constraint also decreases the power consumption of a schedule. Further, applying the techniques proposed in this paper in the context of CD- FGs with fine grained DVS enabled resources under different user constraints remains an interesting research direction.

**AUTHORS**

Rajdeep Mukherjee received the B.Tech degree from the Department of Computer Science and Engineering, University of Calcutta, Kolkata, India, in 2010 and currently pursuing M.S. degree from the Department of Computer Science and Engineering, Indian Institute of Technology, Kharagpur, India. His research interests include verification of VLSI designs, power intent verification, hardware-software co-verification. He is a Research Consultant for Synopsys Inc., USA since 2010.

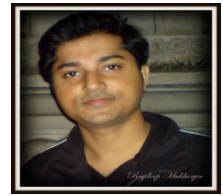

Priyankar Ghosh completed his BE in the year 2003 from the department of Computer Science and Engineering, Jadavpur University and his M.Tech in the year 2006 from the department of Computer Science and Engineering, IIT Kharagpur. Currently, he is pursuing his PhD. degree from the department of Computer Science and Engineering, IIT Kharagpur. His research interests include Artificial Intelligence and Verification of VLSI Designs.

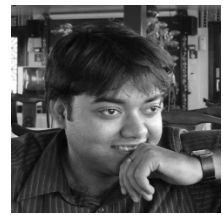

Pallab Dasgupta received the B.Tech., M.Tech., and Ph.D. degrees in computer science from Indian Institute of Technology, Kharagpur, India. He is currently a Professor with the Department of Computer Science and Engineering, IIT Kharagpur. His research interests include Formal verification, artificial intelligence, and VLSI. He has over 100 research papers and two books in these areas. He currently leads the Formal Verification Group, Department of Computer Science and Engineering, IIT Kharagpur (www.facweb.iitkgp.ernet.in/ pallab/forverif.html).

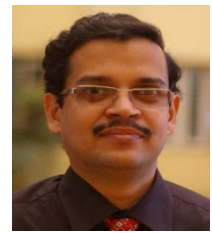

Ajit Pal received the M.Tech. and Ph.D. degrees from the Institute of Radio Physics and Electronics, Calcutta University, West Bengal, India, in 1971 and 1976, respectively. He is presently a Professor with the Department of Computer Science and Engineering, Indian Institute of Technology, Kharag- pur, India. His research interests include real time systems, CAD for VLSI, and computer networks. He has over 90 publications in reputed journals and conference proceedings and a book entitled Microprocessors: Principles and Applications (TMH, 1990).

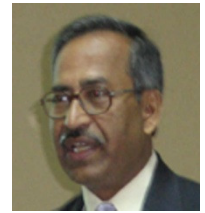